\documentclass[12pt]{iopart}
\usepackage{hyperref}
\usepackage{bbm}
\usepackage{color}
\usepackage{graphicx,epsfig}
\usepackage{amssymb}
\usepackage[normalem]{ulem}

\renewcommand{\tr}{{\rm tr}}
\newcommand{\sini}{\hat{\Pi}_{\rm ini}} 
\newcommand{\sfin}{\hat{\Pi}_{\rm fin}} 
\newcommand{\ps}{\hat{\Pi}_S} 
\newcommand{\pb}{\hat{\Pi}_B} 
\newcommand{\pw}{\hat{\Pi}_W} 
\newcommand{\tw}{\tilde{w}}

\newcommand{\eps}{\epsilon}
\def\be{\begin{eqnarray}}
\def\ee{\end{eqnarray}}
\def\<{\langle}
\def\>{\rangle}
\newcommand{\ini}{{\rm ini}}
\newcommand{\fin}{{\rm fin}}
\newcommand{\diag}{{\rm diag}}
\newcommand{\rk}{{\rm rk}}
\newcommand{\text}{\rm}

\newcommand{\avg}[1]{\big<\big< {#1} \big>\big>}
\newcommand{\savg}[1]{\big< {#1} \big>}

\begin{document}

\title{From single-shot towards general work extraction in a quantum thermodynamic framework \quad}

\author{Jochen Gemmer$^1$ and Janet Anders$^2$}

\address{$^1$ Fachbereich Physik, Universit\"at Osnabr\"uck, Barbarastrasse 7, D-49069 Osnabr\"uck, Germany.}
\address{$^2$ Department of Physics and Astronomy, University of Exeter, Stocker Road, Exeter EX4 4QL, United Kingdom.}
\eads{jgemmer@uos.de, janet@qipc.org}

\begin{abstract} 

This paper considers work extraction from a quantum system to a work storage system (or weight) following reference [1]. An alternative approach is here developed that relies on the comparison of subspace dimensions without a need to introduce thermo-majorisation used previously. Optimal single shot work for processes where a weight transfers from (a) a single energy level to another single energy level is then re-derived. In addition we discuss the final state of the system after work extraction and show that the system typically ends in its thermal state, while there are cases where the system is only close to it. The work of formation in the single level transfer setting [1] is also re-derived. The approach presented now allows the extension of the single shot work concept to work extraction (b) involving multiple final levels of the weight. A key conclusion here is that the single shot work for case (a) is  appropriate only when a \emph{resonance} of a particular energy is required.  When wishing to identify ``work extraction'' with finding the weight in a specific available energy or any higher energy a broadening of the single shot work concept is required. 
As a final contribution we consider transformations of the system that (c) result in general weight state transfers. Introducing a transfer-quantity allows us to formulate minimum requirements for transformations to be at all possible in a thermodynamic framework. We show that choosing the free energy difference of the weight as the transfer-quantity one recovers various single shot results including single level transitions (a), multiple final level transitions (b), and recent results on restricted sets of multi-level to multi-level weight transfers. 

\end{abstract}

\maketitle

\section{Introduction}

The neat characterisation of general classical non-equilibrium processes in terms of fluctuation relations \cite{Evans93,Jarzynski,Crooks,Kawai} has rapidly advanced the general understanding of thermodynamic processes and properties at the mesoscopic scale. Work, in particular, is a thermodynamic quantity of interest and the stochastic fluctuations of work done on a system are captured in the Jarzynski relation \cite{Jarzynski}. The fluctuation relation approach has been extended to quantum systems where probabilistic energy transfers of the system that undergoes unitary evolution are associated with the fluctuating work done on the system  \cite{Tasaki, KURCHAN, Mukamel, Hanggi,CAMPISI2011,Paternostro,Paz}. Again the work for this quantum scenario can be captured in a quantum Jarzynski relation. 
On the other hand, thermodynamic processes for a quantum system can be studied in a setting where the system interacts with a heat bath and a work storage system (or weight) undergoing global unitary dynamics \cite{Janzing,resource,Horodecki2013,Aberg13,SSP14,SSP14,Brandao13b,Skrzypzyk2014,Lostaglio15}. Operation on the system then results in a change of the \emph{work storage system's state} and it is that change that is here associated with ``work''. These approaches are referred to as ``thermodynamic resource theory'' and ``single shot thermodynamics''. Recent papers, e.g. \cite{Horodecki2013, Aberg13,SSP14}, derive upper bounds on the amount of work that can be drawn from a quantum system that starts in a non-equilibrium state in a ``single shot''. 

\begin{figure}[h]
	\begin{center}
	{\includegraphics[width=0.5\textwidth]{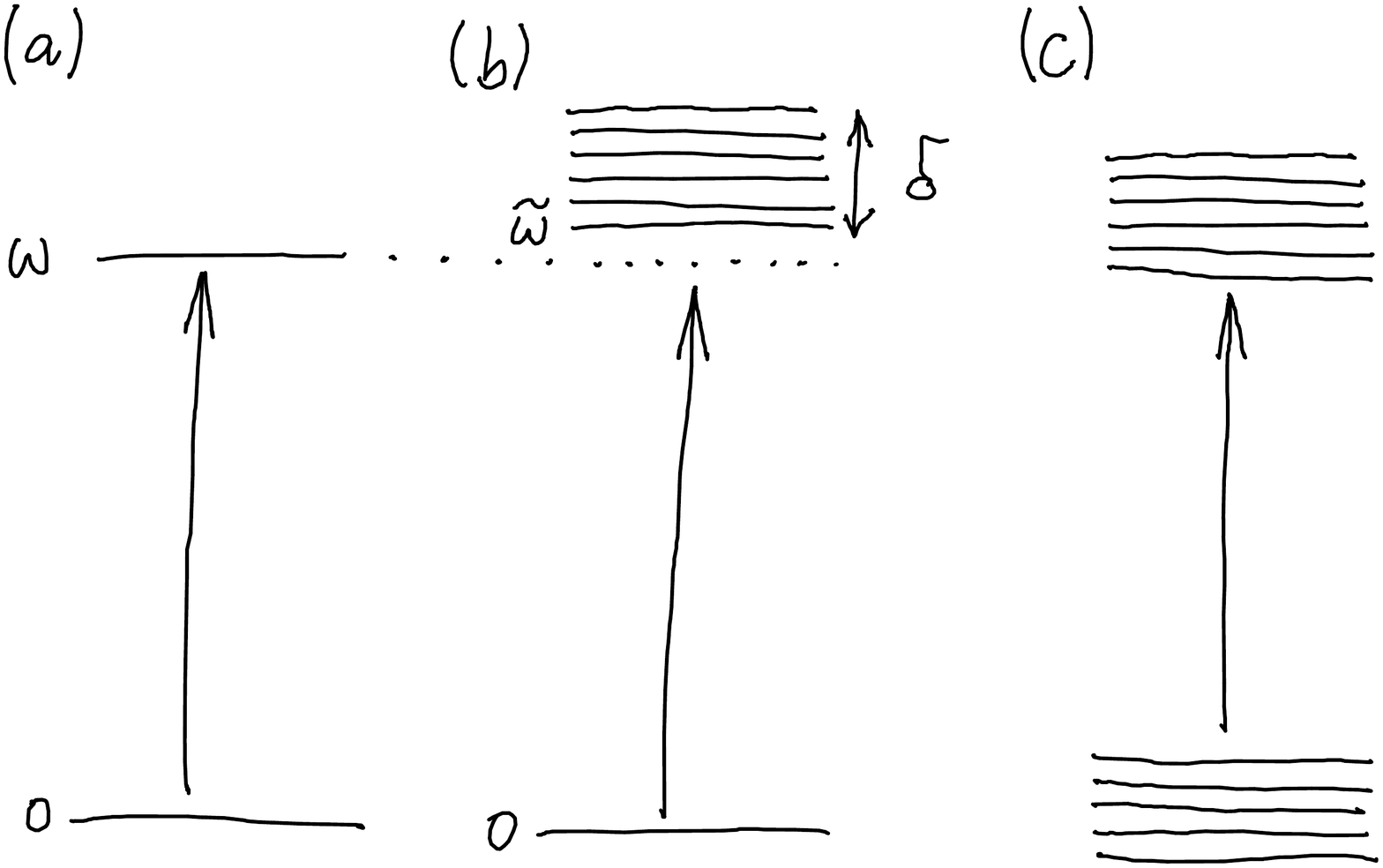}}
\caption{ 
Sketch of the transition of the weight (a) from a single energy eigenstate (of energy $0$) to another single energy eigenstate (of energy $w$), (b) from a single energy eigenstate of energy $0$ to a set of energy eigenstates of energy $[\tw, \tw+\delta]$, and (c) from a range of energy eigenstates to another range of energy eigenstates. 
}
\label{fig:pic}
	\end{center}
\end{figure}

The \emph{single shot work} done by the system is here associated \cite{Horodecki2013} with the transition of the weight from a single energy eigenstate (of energy $0$) to another single energy eigenstate (of energy $w$). This situation is sketched in Fig.~\ref{fig:pic}a. The proof of the bounds provided in \cite{Horodecki2013} relies on established mathematical concepts from quantum information theory and a new majorisation concept called ``thermo-majorisation''. Researchers, in particular those outside of quantum information theory, may find the proof mathematically heavy and are unable to follow the detailed logic. This technical difficulty overshadows the interpretation of the results and may hinder their further development and formulation of experimental tests of single shot results.

Here we aim to re-derive single shot work extraction limits while keeping the technical side as simple as possible. The hope is that stripping the discussion from some of the jargon will allow to focus on the physical meaning of the results, clarify the situation they describe, and develop the argument further to adapt to different physically relevant scenarios. 

The paper is organised as follows. The known work extraction result of \cite{Horodecki2013} for scenario (a) in Fig.~\ref{fig:pic} is re-stated in Section \ref{sec:known} and then re-derived in Section \ref{sec:max}. The final state of the system after this work extraction procedure is explored in Section \ref{sec:fisysta}. Also re-derived, in Section \ref{sec:mwcof}, is the work of formation which was first identified as different from the extractable work in \cite{Horodecki2013}.  The extension of single shot work extraction to general transfer processes is discussed thereafter. Section \ref{sec:disc} is concerned with the extractable work allowing (b) the transition of the weight from a single energy eigenstate of energy $0$ to a set of energy eigenstates of energy $[\tw, \tw+\delta]$. Section \ref{sec:casec} identifies a transfer-quantity that characterises (c) the transition of the weight from an arbitrary range of energy eigenstates to another range of energy eigenstates. Finally, the findings and open 
questions are discussed in Section \ref{sec:discussion}. 

\section{Known results on maximal single shot work extraction}   \label{sec:known}

A key result in single-shot thermodynamics is the identification of a maximal work  \cite{Horodecki2013,Aberg13}
\be \label{eq:ho}
	w^{\max}_\eps =  F^{\min}_\eps (\rho_S) - F(\tau_S),
\ee
that can be extracted with success probability $1 - \eps$ from a system starting in a state $\rho_S$ under so-called \emph{thermal operations} \cite{Horodecki2013}. Here $F(\tau_S) :=-\frac{1}{\beta} \, \ln Z_S$ is the standard free energy associated with the thermal state $\tau_S := \frac{e^{-\beta H_S}}{Z_S}$ for a system Hamiltonian $H_S$ at inverse temperature $\beta$, with $Z_S = \tr[e^{-\beta H_S}]$ the partition function. (In \cite{Horodecki2013} the system is a qubit with the excited state energy tuned such as to have exactly the optimal work value that can be gained, i.e. $H_S = 0 \, |0 \rangle_S \langle 0| + w^{\max}_\eps \, |1 \rangle_S \langle 1|$.) The other quantity, $F^{\min}_\eps (\rho_S)$, is a generalised free energy applicable for the non-equilibrium state $\rho_S$ \cite{Horodecki2013} which will be detailed below. This maximal work is valid for initial states $\rho_S$ that are diagonal in the energy basis and it is at least a lower bound on the maximal work for non-diagonal states \cite{
Horodecki2013, Aberg13, KA15}.  

This result is derived within the thermodynamic resource theory setting 
\cite{Janzing,resource,Horodecki2013,Aberg13,SSP14,SSP14,Brandao13b,Skrzypzyk2014,Lostaglio15} involving three components: the system of interest, $S$, a bath $B$, and a weight (or work storage system) $W$. In the simplest case, the Hamiltonian at the start and the end of the process is assumed to be the same and the sum of the three local terms, $H= H_S + H_B + H_W$. The weight is assumed to have no degeneracies, i.e. all its energy eigenstates have different energies $E_W$. The system's degeneracy is not restricted and we denote the multiplicity of each energy $E_S$ by $M_S(E_S)$ and label each of them by $g_S(E_S) = 1, ..., M_S(E_S)$. The  degenerate bath levels at energy $E_B$ are labeled by $f_B(E_B)$ and their multiplicity is assumed to be exponentially growing with inverse temperature $f_B(E_B) = 1, ..., M_B (E_B) = 1, ..., M_B (0) \, \exp (\beta E_B)$. This exponential growth of the degree of degeneracy (or the density of states) generically results for all systems made of many similar subsystems 
that feature short range interactions \cite{Gemmer2009}. We note that it only holds for a possibly large but finite energy regime. 

\emph{Thermal operations} have been defined \cite{Horodecki2013} as those transformations of the system that can be generated by a
global unitary, $V$, that acts on system, bath and work storage system initially in a product state $\rho_S \otimes \tau_B \otimes |E_W^\ini \>_W\< E_W^\ini|$, with the bath in a thermal state, $\tau_B =  \frac{e^{-\beta H_B}}{Z_B}$, and the work storage system in one of its energy eigenstates, $|E_W^\ini \>$. Conceptually, global unitaries $V$ describe the operation of a ``work extraction machine'' that aims to extract as much energy as possible to the work storage system. Perfect energy conservation is imposed by requiring that the unitary may only induce transitions within an energy shell of total energy $E =E_W+E_S+E_B$. This is equivalent to requiring that $V$ must commute with the sum of the three local Hamiltonians.

The generalised free energy of the non-equilibrium state $\rho_S$ is defined as
\be \label{eq:noneqF}
	F^{\min}_\eps (\rho_S) 
	:=-\frac{1}{\beta} \, \ln \sum_{E_S, g_S} \, e^{-\beta E_S} \, h( E_S, g_S, \eps).
\ee
Here $h(E_S, g_S, \eps)$ is a binary function that determines whether a particular energy eigenstate, $|E_S, g_S\>$, is included in the summation
or not. The value of $h$ depends on $\rho_S$ and on the failure rate $\eps$ that is being accepted for the work extraction. The exact dependence 
will be discussed further in Section \ref{sec:max}. 
The optimal work stated in Eq.~(\ref{eq:ho}) now corresponds to the following task: 
For the weight initially entirely in its ground state (with energy $E_W^\ini=0$) the full system is transferred to a final quantum state such that 
the probability for the weight to be found in an energy eigenstate with energy $E_W^\fin$ is $1-\eps$ where $1>\eps \ge 0$, see Fig.~\ref{fig:pic}a. 
This process is associated with the ``lifting'' of the weight and the energy difference experienced by the weight is identified with 
``extracted work'', $w:= E_W^\fin-E_W^\ini$. For the special case of perfect work extraction, i.e. $\eps =0$, the summation in Eq.~(\ref{eq:noneqF})
includes all energy eigenstates $|E_S, g_S\>$ that are populated in the initial state, i.e. $h( E_S, g_S, 0) = 1$ when $\<E_S, g_S| \rho_S |E_S, g_S\> > 0$, and $0$ otherwise. The maximal work is then 
\be  \label{eq:perfectwork}
	w^{\max}_0 = -\frac{1}{\beta} \, \ln \tr[\tau_S \, \hat{\Pi}_{\rho_S}], 
\ee
where $\hat{\Pi}_{\rho_S}$ is the projector on the support of $\rho_S$.

\section{Work extraction bounds from limits on probability transfer} \label{sec:max}

The derivation of the maximal extractable work presented in \cite{Horodecki2013}  rests on an analysis of state dimensions which is combined with the newly introduced concept of \emph{thermo-majorisation}. This is a variation on \emph{majorisation}, an important tool in the study of doubly stochastic matrices \cite{vonNeumann,Ando,Mat} and quantum channels in particular \cite{Ref1, Ref2}. Here we present a derivation of the same result without a need of invoking thermo-majorisation explicitely. We hope this re-derivation is more straightforward to follow for researchers wishing to familiarise themselves with single-shot thermodynamics. 

The rational of the presented approach is to compare dimensions of the involved projective subspaces to conclude what transformations on the system are possible in the setting given and what maximal work extraction they enable. The re-analysis opens avenues of generalising the previous result, valid for transitions of the weight from single energy level to single energy level,  to processes where the weight transfers between multiple energy levels, which we will explore in Sections  \ref{sec:fisysta} and \ref{sec:disc}. The work required to form a non-equilibrium state is also described, in Section \ref{sec:mwcof}. 

The desired  thermal transformation is of the form 
\be \label{eq:map}
 	\eta : = \rho_{SB} \otimes |0\>_W\<0| \stackrel{V}{\longrightarrow} \sigma_{SB} \otimes |w\>_W\<w| =: \eta'(V),
\ee
where the system and bath start in a fixed state $\rho_{SB}$ and end in some state $\sigma_{SB}$, while the work storage system lifts from a single energy level, $|0\>$, to another single energy level, $|w \>$, by an energy $w>0$, see Fig.~\ref{fig:pic}a. Such a transformation, however, turns out to be impossible for a large class of initial states, $\rho_{SB}$, namely those which have full rank. Thus one allows transformations under which the l.h.s. of (\ref{eq:map}) is transformed into the r.h.s. only up to a success probability $1-\epsilon$, as already introduced in Sect.~\ref{sec:known}. 

Since the global unitary, $V$, is fully energy conserving we may treat the global dynamics for each global energy shell $E$ separately and later combine their contributions. Note that those energy shells are projective subspaces and so are the individual energy levels that make up the shell. We denote the local projectors defined by their associated local energies $E_S, E_B$ and $E_W$ by $\ps(E_S), \pb(E_B)$ and $\pw(E_W)$. 

To find the maximal work $w^{\max}_\eps$ we first consider a probability $P^E_{\ini} \le 1$ which falls initially into a projective subspace $\sini^E$ of dimension $d^E_{\ini}$, i.e. $P^E_{\ini}= \tr [ \eta \, \sini^E]$ for the initial state $\eta$. The key task here is to decide if this probability can be transferred \emph{entirely} into a projective subspace $\sfin^E$ under unitary transformations $V$. This is of course possible if
\be \label{eq:cond1}
	\tr[ \sfin^E \, V \, \sini^E \, \eta \, \sini^E \, V^{\dag}]
	\stackrel{!}{=} \tr[ \sini^E \, \eta \, \sini^E] = P^E_{\ini}
\ee
holds. Rewriting this condition using the non-trivial eigenstates $|n\rangle $ of $V \, \sini^E \, \eta \, \sini^E \, V^{\dag}$, i.e. $V \, \sini^E \, \eta \, \sini^E \, V^{\dag} |n\> = p_n |n\>$ with $p_n > 0$ for $n = 1, ..., d^E_{\ini}$,
implies
\be 
	\sum_{n=1}^{d^E_{\ini}} p_n  \, \<n | \sfin^E |n\>
	\stackrel{!}{=} \ \sum_{n=1}^{d^E_{\ini}} p_n = P^E_{\ini}
\ee
and therefore one must have $ \<n | \sfin^E |n\> = 1$ for $n = 1, ..., d^E_{\ini}$. This is only possible if the dimension of the space into which the probability is mapped, $d^E_\fin := \rk[\sfin^E]$, is at least  the same as the initial dimension, i.e. the condition becomes
\be  \label{eq:cond2}
	d^E_\fin  \ge d^E_\ini.
\ee
If not the whole probability $P^E_{\ini}$ but a slightly reduced probability $(1-\eps) P^E_{\ini}$ is to be transferred, then one may replace $\sini^E$ in Eq.~(\ref{eq:cond1}) with any projector $\sini^E (\eps)$ that fulfils 
\be \label{eq:projeps}
	\tr [ \sini^E (\eps) \, \eta ] \geq P^E_{\ini}(1-\eps).
\ee
Defining the smallest initial subspace dimension as $d^E_{\ini}(\eps, \eta) := \min_{\sini^E (\eps)} \, \dim( \sini^E (\eps)) \le d^E_{\ini}$, the final space dimension condition, when allowing a small error probability $\eps$ in the transfer, relaxes to
\be  \label{eq:cond2eps}
	d^E_\fin  \ge d^E_\ini (\eps, \eta).
\ee
This constraint is now the key condition to establish what the maximum extractable work is for the thermal operations specified in Eq.~(\ref{eq:map}). 

\begin{figure}[t]
	\begin{center}
	\includegraphics[width=0.58\textwidth]{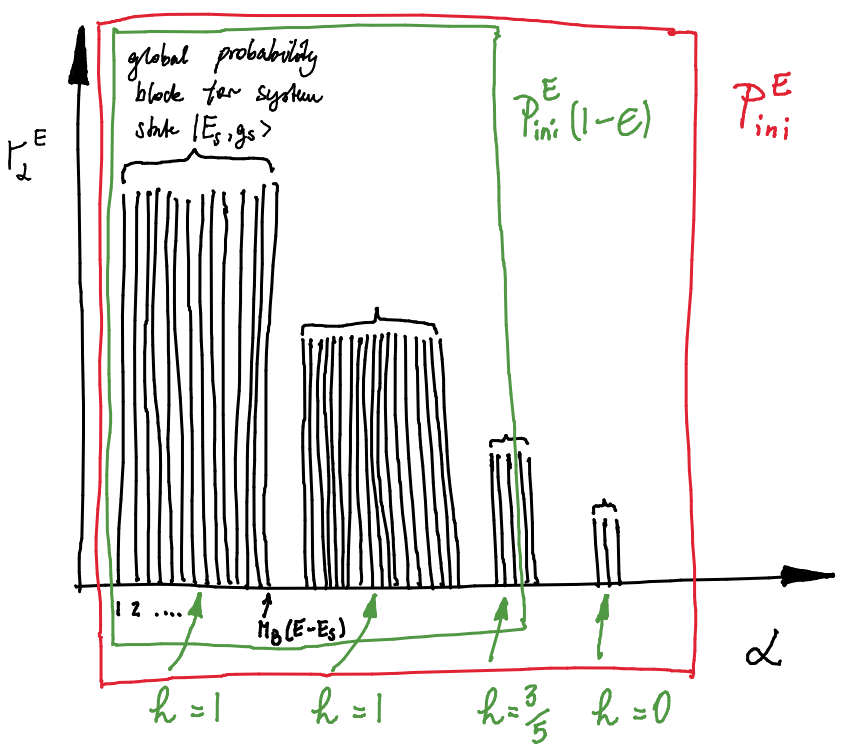}
        \caption{\label{fig:blocks} 
       Eigenvalues $r^E_{\alpha}$ of global initial state $\eta$ in energy shell $E$ given in Eq.~(\ref{eq:eigen}) arranged in decreasing order by their index $\alpha$. For any particular system state $|E_S, g_S\>$ there is a whole block of global state eigenvalues of the same magnitude that arises due to the bath's degeneracy $M_B (E-E_S)$ at energy $E_B = E-E_S$. The full initial probability, $P^E_\ini$, is indicated by the red box. The slightly reduced probability that is to be transferred, $P^E_\ini \, (1-\eps)$, is indicated by the green box. To make up the slightly reduced probability blocks of probabilities are either fully included ($h=1$), fractionally included (e.g. $h=3/5$), or not included ($h=0$) in the summations Eq.~(\ref{eq:dinieps}) and Eq.~(\ref{eq:maxw}).    
        }
	\end{center}
\end{figure}

Since $V$ cannot mix energy shells, it is necessary to check the condition for each energy shell $E$. We interpret the projector $\sfin^E$ used above as the operator that projects on the non-trivially populated subspace of the final global state, $\sigma_{SB} \otimes |w\>_W\<w|$, with energy $E$. 
By construction one has the relation
\be \label{eq:pfine}
	\sfin^E = \sum_{E_S} \ps (E_S) \otimes \pb (E-E_S-w) \otimes \pw (w).
\ee
We can now identify the dimension of $\sfin^E$, 
\be \label{eq:dfine}
	d^E_{\fin}
	&=&\sum_{E_S} M_S(E_S) \, M_B(E-E_S-w) 
	= \sum_{E_S} M_S(E_S) \, M_B(E) \, e^{-\beta E_S-\beta w} \\
	&=& M_B(E) \, e^{-\beta w}  \, Z_S,
\ee
where $Z_S =  \sum_{E_S} M_S(E_S) \, e^{-\beta E_S}$ is the thermal equilibrium partition function at
inverse temperature $\beta$ for the system. The second step is to consider the initial subspace projectors in the energy shell $E$,
\be \label{eq:pinie}
	\sini^E= \sum_{E_S} \ps (E_S) \otimes \pb (E-E_S) \otimes \pw (0).
\ee
In the following we aim to specify the dimension $d^E_{\ini} (\eps, \eta)$ of a sub-projector of $\sini^E$ in whose associated subspace lives the fraction $1-\eps$ of the initial state ($\eta$) population ($P^E_{\ini}$) in energy shell $E$.
For $\eps =0$ and $\rho_{SB}$ a full rank state the dimension of $\sini^E$ itself is just
\be \label{eq:dinie}
	d^E_{\ini} (0)
	= \sum_{E_S} M_S(E_S) \, M_B(E-E_S) 
	= M_B(E)  \, Z_S.
\ee
However, to determine the initial dimension $d^E_{\ini}(\eps, \eta)$ for a finite failure rate $\eps >0$ we will need to specify the global initial state $\eta$ a little more. While not the only choice, here we consider the class of initial global product states previously discussed \cite{Horodecki2013, Aberg13, SSP14, Skrzypzyk2014},
\be \label{eq:roini}
	\eta = \rho_S \otimes \tau_B  \otimes |0\>_W\<0|
\ee
where $\rho_S$ is a system state that is diagonal in the basis of the system Hamiltonian and the bath is in a thermal state $\tau_B$. We will label the eigenvalues of $\rho_S$ when diagonalised in the system energy eigenbasis by $\lambda (E_S, g_S)$ where $E_S$ is the corresponding energy and $g_S$ the degeneracy index. By construction all \emph{non-zero} eigenvalues of $\eta$ in the energy shell $E$, i.e. the non-zero eigenvalues of $\sini^E \, \eta \, \sini^E$, are then
\be \label{eq:eigen}
	r^E(E_S, g_S, f_B) = \lambda (E_S, g_S) \, \frac{e^{-\beta(E-E_S)}}{Z_B}. 
\ee
Here $f_B = 1, ..., M_B$ is the degeneracy index of the bath at energy $E_B = E-E_S$; but it does not affect the magnitude of $\eta$'s eigenvalues. The eigenvalues thus have a high multiplicity, given by the bath multiplicity $M_B$. The eigenvalues can be re-labeled with  
$\alpha = 1, ..., d^E_{\ini}(0)$ as $r^E (E_S, g_S, f_B) = r^E_{\alpha}$, such that they are arranged in decreasing order, $r^E_{1} \ge r^E_2 \ge r^E_3 ... $, see Fig.~\ref{fig:blocks}. The spectra of different energy shells $E$, as sketched in Fig.~\ref{fig:blocks}, differ only by a factor of $e^{-\beta E}$ for the individual values while their multiplicity $M_B$ differs by a factor of $e^{\beta E}$. The population probability of the global state $\eta$ in energy shell $E$ is given by $\sum_{\alpha =1}^{d^E_{\ini}(0)}  r^E_{\alpha} = P^E_{\ini}$.

Now we are ready to identify the dimension $d^E_{\ini}(\eps, \eta)$ of the subspace within the energy shell $E$ where most of the initial probability, $P^E_{\ini} \, (1-\eps)$, lives. In the sum above one simply has to add up 
the $\alpha$ to $d^E_{\ini}(\eps, \eta) \le d^E_{\ini}(0)$ which must be chosen such that the probability reduces to the desired level,
\be \label{eq:dint}
	\sum_{\alpha=1}^{d^E_{\ini}(\eps, \eta)}  r^E_{\alpha}  \ge  P^E_{\ini} \, (1-\eps).
\ee

Comparing with Eq.~(\ref{eq:dinie}) the final dimension is given by 
\be \label{eq:dinieps}
	d^E_{\ini} (\eps, \eta) = \sum_{E_S, g_S}  M_B (E-E_S)  \, h(E_S, g_S, \eps)
\ee
where $h(E_S, g_S, \eps)$ determines which terms are included in the sum, as visualised and explained in Fig.~\ref{fig:blocks}. An $h$-value of 1 (0) indicates that a whole block of global eigenvalues all corresponding to system state $|E_S, g_S\>$ is (not) included in the summation, while a fractional $h$-value indicates that a fraction of the eigenvalues in a block is included, see Fig.~\ref{fig:blocks}. 

The value of $d^E_{\ini} (\eps, \eta)$ must be chosen as the smallest \emph{integer} that fulfills (\ref{eq:dint}). It can happen that $\eps$ has a value that would require to split up an individual eigenvalue $r^E_{\alpha}$ in order to fulfill (\ref{eq:dint}) as an equality, i.e. $\sum_{\alpha=1}^{n}  r^E_{\alpha} +  a \, r^E_{n+1}= P^E_{\ini} \, (1-\eps)$  where $n$ is an integer and $0<a<1$ a real fraction. In this case the eigenvalue $r^E_{n+1}$ must be fully included, i.e. $h$ must be chosen as the larger proper fraction available. For example, having to split the 3rd of 5 eigenvalues in a block $(E_S, g_S)$ the associated $h(E_S, g_S, \eps)$ must be chosen 3/5 rather than 2/5, in order to guarantee that the failure probability is strictly less than $\eps$. 

For a map that is identical to (\ref{eq:map}) up to a failure probability $\eps$ we have obtained the initial and final subspace dimensions for each energy shell $E$. They allow us to check the maximal work extraction condition Eq.~(\ref{eq:cond2eps}) for each $E$. One has 
\be \label{eq:maxw}
\fl	d^E_{\fin}  =  M_B(E) \, e^{-\beta w}  \, Z_S 
	& \phantom{w} \ge&  \sum_{E_S, g_S}  M_B (E) \, e^{- \beta E_S} \, h(E_S, g_S, \eps)  = d^E_{\ini} (\eps, \eta)  \\
\fl	&w \le& - \frac{1}{\beta} \, \ln \sum_{E_S, g_S}  {e^{- \beta E_S}} \, h(E_S, g_S, \eps) - F(\tau) = w^{\max}_\eps. \nonumber
\ee
The last line shows that because of the trivial dependence of $d^E_{\ini}(\eps, \eta)$ on $E$, the latter drops out and the expression for the maximal extractable work stated in Eq.~(\ref{eq:ho}) is recovered \cite{Horodecki2013}. Note, that the definition of $h$ has been extended from a binary function, used in Eq.~(\ref{eq:ho}), to include rational fractions. The optimal work described by Eq.~(\ref{eq:maxw}) is thus marginally larger than the work described by Eq.~(\ref{eq:ho}). To recap, this work is maximal under variation of the global unitaries $V$ that realise the map (\ref{eq:map}). The bound is tight when $\eps$ is chosen such that no splitting of eigenvalues would be necessary. 

Note, that using the eigenvalues of the thermal state, $\tau_S$, denoted by $t(E_S) := \frac{e^{- \beta E_S}}{Z_S}$, the maximal extractable work (\ref{eq:ho}) can also be written as 
\be \label{eq:maxwork2}
	w^{\max}_\eps 
	=  -\frac{1}{\beta} \,  \ln \sum_{E_S, g_S} \, t(E_S) \, h( E_S, g_S, \eps).
\ee
Clearly, if $h$ was 1 for all its arguments, i.e. all the eigenvalues in Fig.~\ref{fig:blocks}  have to be included, then the maximal extractable work is 0. For $\eps = 0$ no probability can be lost and $h=1$ for all energy levels with non-zero population. Work can thus only be extracted if the initial state has rank less than the system Hilbertspace dimension or $\eps >0$ or both.
 
\section{The final system state} \label{sec:fisysta}

The work extraction protocol, Eq.~(\ref{eq:map}),  has mapped $\eta \to \eta' (V)$. It is natural to ask in what reduced state $\sigma_S = \tr_{BW}[\eta' (V)]$ the system is left as a result. We will here answer this question.

A crucial point to note is that not a single, unique unitary $V$ corresponds to maximum work extraction but that there are infinitely many. Any unitary that maps the projective energy subspaces $\sini^E (\eps)$, with dimensions $d^E_{\ini}(\eps, \eta)$ calculated in Section \ref{sec:max}, onto the  projective subspaces  $\sfin^E$ will give rise to maximum work extraction. The large set of unitaries with this property forms a subset of all allowed unitaries, i.e. those that commute with $H$. As a first step we look at the average final state, obtained through the application of each of these unitaries and integration over the respective Haar measure, which we denote by $\avg{\cdot}$. Details on integration over Haar unitaries can be found e.g. in \cite{Goldstein2010}. Note that even though the unitaries $V$ commute with $H$, $\eta' (V)$ itself need not be diagonal w.r.t. to the product basis of local energy eigenstates that defines the energy shells. While $H$ and $\eta' (V)$ share a simultaneous eigenbasis there is no guarantee, due to $H$ having degeneracies, that this is the product basis of local energy eigenstates and thus $\eta' (V)$ can have off-diagonal elements in the degenerate blocks of $H$. 
For the diagonal part
of $\eta' (V)$ w.r.t. the above basis consisting of products of energy eigenstates  of individual subsystems, $\eta'_{\diag} (V)$, one finds
\be  \label{eq:find}
	\avg{\eta'_{\diag}} \propto e^{-\beta H_S} \otimes  e^{-\beta H_B}   \otimes |w\>_W\<w|,
\ee
see \ref{derfin} for details of the derivation.
This is a factorized state where the system and bath are each in thermal states at inverse temperature $\beta$. However, this result in itself is not sufficient to conclude that typically an almost thermal state of the system results from an individual  $V$. One needs to calculate also the relative variances of $P_{\fin} (E, E_S, g_S)$. The latter are the portions of the diagonal elements at $E_S, g_S$ of the final reduced state of the system $\sigma_S$ that correspond to the parts of the total final state living in energy shell $E$, i.e., $\< E_S, g_S|\sigma_S|E_S, g_S\> = \sum_E P_{\fin} (E, E_S, g_S)$. Hence, if these relative fluctuations are small one will typically get diagonal elements of $\sigma_S$ that resemble their averages, which in turn coincide, according to Eq.~(\ref{eq:find}), with those of a thermal state.
Concretely we obtain for those relative fluctuations, see \ref{derfin}, 
\be \label{eq:fluc}
	\frac{\Delta P_{\fin} (E, E_S, g_S)}{\avg{P_{\fin} (E, E_S, g_S)}} \approx \frac{1}{\sqrt{M_B(E - E_S - w)}},
\ee
which is indeed very small due to the large multiplicity of the bath $M_B(E - E_S - w) = M_B(E) \, e^{-\beta (E_S + w)}$ 
for energies $E >> E_S + w$. Off-diagonal elements are considered separately, as detailed in \ref{derfin}, but it turns out that the variance to their vanishing mean also scales as $\frac{1}{\sqrt{M_B(E - E_S - w)}}$. Thus, for a unitary drawn at random all elements of a typical final system state $\sigma_S$ are $\frac{1}{\sqrt{M_B(E - E_S - w)}}$ close to the average thermal state if the bath is large. The entire above reasoning is along the lines of what has become known as ``quantum typicality'' \cite{Reimann2007, Goldstein2006, Popescu2006, Gemmer2003}.

Having realised that optimal work extraction can result in a multitude of final states, with typical  state being close to the 
thermal state, one may wonder why it is not possible to extract more work in those rare instances where the final state happens to be fairly different from the thermal state. Here it is important to note that the final state need not to be factorized between system and bath (c.f. Eq.~(\ref{eq:find}) only describes the diagonal part). Indeed, it is known  \cite{Lubkin1978, Page1993, Lloyd1988, Sen1996, Gemmer2001} that only a negligible set of the final states will be factorized, especially  when the degeneracies of the bath $B$ are large. However, to repeat the same work extraction process requires an initially fully factorized state, thus preventing repetition and further work extraction along the same route for the majority cases. 

Finally, a concern may be that one could introduce a ``fresh bath''  that is uncorrelated with the final system in order to extract even more work from a non-thermal final system state. Here it is important to realise that the work extraction process with the first bath left the system not necessarily in a thermal state, but certainly in a state with full rank, $\sigma_S$. This is due to the fact that maximum work extraction requires a non-zero population of every dimension spanned by $\sfin^E$, see Eq.~(\ref{eq:pfine}). As can be seen from Eq.~(\ref{eq:perfectwork}) a full rank system state means that there is no potential for further ``perfect'' work extraction, i.e. moving weight population from $|0\>_W \to |w\>_W$ with certainty ($\epsilon = 0$), even if new baths are brought in. If the first work extraction process is imperfect ($\epsilon \neq 0$), i.e. the weight ends with population in $|0\>_W$ and $|w\>_W$, in general weight and system will be correlated, thus also violating the initial factorisation condition, see (\ref{eq:map}), for a second work extraction process. Nevertheless, for completeness, suppose that the result factorises and the weight is chosen such that it has additional energy levels $|w_2\>$ and $|w+w_2\>$, where $w_2$ is adjusted to match $w_2 = w^{\max}_{\eps} (\sigma_S)$ of Eq.~(\ref{eq:maxwork2}) applied to $\sigma_S$. One can now indeed perform a second imperfect work extraction which may very well increase the mean energy in the weight system. However, this will lead to a weight level population of at least three of the four eigenlevels, $|0\>_W, |w\>_W, |w_2\>_W, |w+w_2\>_W$. This scenario is outside of what is allowed in case (a), see Fig.~\ref{fig:pic}a, but will be discussed as case (b) in Section \ref{sec:disc}. 
If, on the other hand the first work extraction process is a perfect one, weight and system necessarily factorise and a second, imperfect, work extraction can be added. This will result in overall imperfect transfer of $|0 \>$ to $| w^{\max}_{\eps} (\rho_S)\>_W$ as a result of the total process instead of raising perfectly to $| w^{\max}_0 (\rho_S)\>_W$ in the first step alone. So one could have obtained the same result if one had allowed for imperfect work extraction in the first place since an increase of $\epsilon$ always allows for an increase of work, $w^{\max}_{\eps} (\rho_S) >w^{\max}_0 (\rho_S)$. This shows that there is no second law conflict with a final state being non-thermal.
  
\section{Minimal work cost of formation} \label{sec:mwcof}

A key finding of the single shot thermodynamics approach presented in \cite{Horodecki2013} is the realisation that, in analogy to entanglement of formation and distillation, the work required to form a non-equilibrium state may differ from the work that can be extracted from that state. Here we will now re-derive the minimal cost of formation, $w^{\min}$, of engineering with certainty ($\eps =0$) a diagonal state $\sigma_S$ from a thermal state $\tau_S = \frac{e^{-\beta H_S}}{Z_S}$, under thermal operations. 

Considering again global unitaries $V$ that commute with the Hamiltonian, the desired operation is
\be \label{eq:formationmap} 
	 \eta := \tau_S \otimes \tau_B \otimes |w\>_W\<w| \stackrel{V}{\longrightarrow} \sigma_{SB} \otimes |0\>_W\<0| :=
	 \eta',
\ee
with any final state permitted such that $\tr_B[\sigma_{SB}] = \sigma_S$. 
Using the same reasoning that lead to Eq.~(\ref{eq:eigen}) the eigenvalues $r^E(E_S, g_S, f_B, E_W)$  of the global initial state $\eta$ in a particular energy shell $E$ are given by
\be \label{eq:eigenform} 
	 r^E(E_S, g_S, f_B, E_W)= \frac{e^{- \beta (E - w)}}{Z_S \, Z_B} \, \delta_{E_W,w},
\ee
where $g_S$ and $f_B$ are the degeneracy indices labelling the various energy eigenstates for $E_S$ and $E_B=  E- E_S - E_W$, respectively. By construction the global eigenvalues are only non-zero when the weight has energy $E_W = w$ and the non-zero eigenvalues in shell $E$ are all equal. 

The eigenvalues of the desired final system state $\sigma_S$ which is assumed diagonal in the system's energy eigenbasis $\{|E_S, g_S\>\}$ of $H_S$ are denoted by $s(E_S, g_S) = \tr[ \sigma_{SB} \, |E_S, g_S\>\<E_S, g_S|]$. The contributions stemming from the energy shell $E$ to the diagonal elements of the final system state $\sigma_S$ may thus be written as
\be \label{eq:redsum} 
\fl	s^E(E_S, g_S) = \sum_{f_B(E-E_S)}  \<0|  \<E-E_S, f_B| \<E_S, g_S| \, \eta' \, |E_S, g_S\>  |E-E_S, f_B\> |0\>,
\ee 
where the sum is performed over all bath degeneracy indices at energy $E_B = E-E_S$. We note that since the initial state has no off-diagonal elements between different system energies, say, $E_S$ and $E_S'$, and the unitary $V$ commutes with $H$ no such off-diagonal elements can be generated under $V$.  There may be off-diagonal elements within a single energy subspace $E_S$, i.e. $\<E_S, g_S| \sigma_{S} \, |E_S, g'_S\> \not = 0$. These off-diagonal elements can always be ``avoided'' though by choosing the basis $\{|E_S, g_S\>\}$ for the system subspace of energy $E_S$ appropriately. Thus, w.l.o.g. the state $\sigma_S$ can be assumed diagonal with diagonal elements as given in (\ref{eq:redsum}).

Since  $V$ is unitary, not only the eigenvalues of the initial state $\eta$ (in energy shell $E$) but also those of the final state $\eta'$ (in energy shell $E$) are given by Eq.~(\ref{eq:eigenform}), i.e. they are either $0$ or $\frac{e^{- \beta (E-w)}}{Z_S \, Z_B}$. The summands on the r.h.s. of (\ref{eq:redsum}) are just expectation values of the state $\eta'$ and since any expectation value of a state is upper bounded by the largest eigenvalue of the respective state we find from (\ref{eq:redsum})
\be \label{eq:einzelval}
\fl	s^E(E_S, g_S) \leq \sum_{f_B(E-E_S)}  \frac{e^{- \beta (E-w)}}{Z_S \, Z_B} 
	= \frac{e^{- \beta (E-w)}}{Z_S \, Z_B} \, M_B(E-E_S)=\frac{e^{- \beta (E_S-w)}}{Z_S}\frac{ M_B(E)e^{- \beta E}}{Z_B}.
\ee
The probability of the final state $\sigma_S$ to be found in state $|E_S, g_S\>$, denoted by $s(E_S, g_S)$, is then obtained by summing over
contributions from all the energy shells, 
\be \label{eq:sumval}
	s(E_S, g_S)=\sum_E s^E(E_S, g_S) \leq \frac{e^{- \beta (E_S-w)}}{{Z_S}},
\ee
where $\sum_E  M_B(E)e^{- \beta E} = Z_B$ was used. Importantly this inequality must be fulfilled for all pairs $E_S, g_S$.
Using the eigenvalues of the thermal state of the system and rearranging one obtains
\be 
	\forall E_S, g_S: w \geq \frac{1}{\beta} \, \ln \frac{s(E_S, g_S)}{t(E_S)}
\ee
as the condition on the work cost of formation. 
One can see now that the energy eigenstate $|E^*_S, g^*_S\>$ with the largest ratio $\frac{s(E^*_S, g^*_S)}{t(E^*_S)} = \max_{E_S, g_S} \frac{s(E_S, g_S)}{t(E_S)} =: \mu$ is the one that constraints the whole transformation. The tight lower bound on the work cost of formation is then
\be \label{eq:wcostbound}
	w \geq \frac{1}{\beta} \, \ln \mu = : w^{\min}.
\ee
$\mu$ can also be expressed as 
$\mu = \min \{\lambda: s(E_S, g_S) \le \lambda \, t(E_S) \, \forall E_S, g_S \}$ or 
\be \label{eq:workcost}
	\mu = \min \{\lambda: \sigma_S \le \lambda \, \tau_S \}
\ee
for $\sigma_S$ the desired diagonal state of the system. 
The work cost Eq.~(\ref{eq:wcostbound}) for the ideal map Eq.~(\ref{eq:formationmap}) can easily be extended to allow final 
states in an $\eps$ vicinity of the desired state $\sigma_S$. $\mu$ is then replaced by 
$\mu^{\eps} = \inf_{\sigma_S^{\eps}} \, \min \{\lambda: \sigma_S^{\eps} \le \lambda \, \tau_S \}$ where $\sigma_S^{\eps}$ 
is close to $\sigma_S$, $||\sigma_S^{\eps} - \sigma_S|| \le \eps$, where $|| \cdot ||$ is the trace norm \cite{Mat}.

\section{Generalisation of work extraction to multiple levels in the final state of the  work storage system} \label{sec:disc}

We have seen in the previous section that the single shot concept gives sensible results for work extraction when single levels of the weight are 
being considered (case (a) in Fig. 1). While for truly microsopic systems this may be a suitable analysis, case (a) is not the right picture for 
mesoscopic or macroscopic work extraction process. It is not feasible to lift a work storage system from a true quantum energy eigenstate to another 
true energy eigenstates. The density of states is so large that it will be practically impossible to pick a single eigenstate of a macroscopic 
system \cite{Reimann2007}. This indicates that an extension of the single shot thermodynamic framework to transitions between multiple 
levels, see Fig.~\ref{fig:pic}b and ~\ref{fig:pic}c, is desirable. 

The approach presented above now allows a straightforward generalization to thermal processes where the work storage system is transferred from the 
ground state to not just a single energy level, but an interval $\delta$ of energy levels, see Fig.~\ref{fig:pic}b. To do so we will now turn to a 
harmonic oscillator work storage system  with energy spacing $\Delta E$, similar to the one considered in \cite{Skrzypzyk2014}, and differing from 
the qubit work storage system used in \cite{Horodecki2013}. For simplicity we will assume that also the system and the bath have 
oscillator-like equidistant energy eigenlevels of integer multiples of $\Delta E$. Again, the weight shall have no degeneracies whatsoever 
while the system and bath again have degeneracies $M_S(E_S)$ and $M_B(E_B)$ at their energies $E_S$ and $E_B$, respectively.  

Work extraction of an amount of work $\tw_\eps$ is now identified with the transition (with probability $1-\epsilon$) of the work storage 
system from its ground state to any state living in the energy subspace $[\tw_\eps , \tw_\eps +\delta]$, see Fig.~\ref{fig:pic}b. The ideal 
transfer ($\eps =0$) is now associated with the map 
\be \label{eq:mapb}
	\eta: = \rho_S \otimes \tau_B \otimes |0\>_W\<0| \stackrel{V}{\longrightarrow} \sigma_{SBW}
\ee where the final state of the weight $\sigma_W := \tr_{SB}[\sigma_{SBW}]$ is now mixed with support in the interval specified. Here the general case of entangled states between weight and the rest is allowed. If the role of these correlations is of interest, then restrictions to product states could be considered.

To account for the different (larger) space of possible final states one simply has to restart the calculation of $\sfin^E$ and $d^E_{\fin}$, as detailed in Sec.~\ref{sec:max}, and enlarge the final  Hilbert space in order to comprise all final global states with the weight in the specified local energy subspace. This ``new'' final  Hilbert space can be described as a sum over many ``old'' ones. Restarting from  (\ref{eq:dfine}) one can introduce a pertinent summation over weight energy levels so that the new $d^E_{\fin}$ reads:
\be \label{dfinesum}
	d^E_{\fin} =M_B(E) \, Z_S \sum_{E_W=\tw}^{\tw+ \delta} e^{-\beta E_W}.
\ee
The sum on the r.h.s. now depends on the level spacing $\Delta E$ which did not play any  role so far,
\be \label{sums}
	\sum_{E_W=\tw}^{\tw+ \delta} e^{-\beta E_W}  
	= \sum_{n=\tw/\Delta E}^{(\tw+ \delta)/\Delta E} e^{-\beta \Delta E \, n}
	= \frac{e^{-\beta \tw}(1-e^{-\beta (\delta + \Delta E) })}{1-e^{-\beta \Delta E }},
\ee
where $n$ is a natural number that labels the eigenvalues in the allowed final interval. Substituting this for $e^{-\beta w}$ in (\ref{eq:dfine}) then leads to
\be
 	\tw^{\max}_{\eps} = w^{\max}_{\eps} + {1 \over \beta} \, \ln \frac{1 - e^{- \beta (\delta + \Delta E) }}{1-e^{-\beta \Delta E }}.
\ee
Increasing the width $\delta$ of the allowed final energy regime for the weight beyond a single energy level ($\delta > 0$) leads to an additional term with respect to the single final energy level situation characterised by Eq.~(\ref{eq:ho}). For large energy intervals ($\delta$) and small energy spacings ($\Delta E$) the last expression quickly converges to
\be \label{eq:hoa}
 	\tw^{\max}_{\eps} = w^{\max}_{\eps} - \frac{\ln(\beta \Delta E)}{\beta}.
\ee
Expression (\ref{eq:hoa}) holds under the conditions that compared to the ``thermal energy'', $\frac{1}{\beta}$ the final energy range is large, $\delta >> \frac{1}{\beta}$ and the energy spacing  of the work storage system, $\Delta E$, is small $\Delta E << \frac{1}{\beta}$. Physically this means that the final energy range $\delta$ has to comprise a large number of energy eigenstates of the work storage system. 

We note that the extractable work $\tw^{\max}_{\eps}$ for case (b) in Fig.~\ref{fig:pic} is \emph{increased} in comparison to the extractable work $w^{\max}_{\eps}$ of case (a) in the limit considered ($\Delta E << 1/\beta$). The surplus on the r.h.s. does not depend on $\delta$ as long as the above conditions are fulfilled. However, the work does vary with the level spacing and can go up to infinity for $\Delta E \to 0$.

To judge whether these conditions are fulfilled for meso/macro-scopic work storage systems, it is instructive to consider realistic numbers.
At room temperature the thermal energy is $\frac{1}{\beta} \approx 2.5 \cdot 10^{-2}$~eV; about the energy of a single optical phonon in a solid. 
Thus allowing  a final energy range of $\delta$ for the macroscopic weight system that is large compared to the energy of a single phonon, 
condition $\delta >> \frac{1}{\beta}$ is fulfilled. For the work storage system one can imagine a pendulum with eigenfrequency 1~Hz which results 
in an energetic level spacing of $\Delta E \approx 4\cdot 10^{-15}$~eV$<< \frac{1}{\beta}$ thus fulfilling the second condition. Level spacings of
many meso/macro-scopic work storage systems, such as batteries, are even smaller by several orders of magnitude. These example numbers indicate that 
both conditions will generally hold for meso/macro-scopic systems.

Maybe surprisingly the additional term, $-\ln(\beta \Delta E)/\beta$, is an up-shift as sketched in Fig.~\ref{fig:pic}b, implying a work above that of Eq.~(\ref{eq:ho}). This seems to indicate a conflict of Eq.~(\ref{eq:hoa}) with the second law, but this is only a problem in a naive sense. To properly discuss second law violations one would have to build a cyclic machine that enables work extraction. It is possible to close  the desired work extraction transformation,  Eq.~(\ref{eq:mapb}), to a cycle, however, it is not possible to run the cycle again. This is because the weight started in an energy eigenstate, $|0\>_W$, while ending in a reduced state $\sigma_W$ which is \emph{not} an energy eigenstate. Therefore there is no true second law violation as the process cannot be repeated using the same weight. Exploring if there is a link between the apparent violation of  the second law in Eq.~(\ref{eq:hoa}) and the existence of weight-system or system-bath correlations \cite{Marti14,HSAL11,Esposito10} in the final state in Eq.~(\ref{eq:mapb}) is an interesting future avenue. 

Finally, the implication of Eq.~(\ref{eq:hoa}) is that one can extract \emph{more work} when  the final weight state is allowed to live in a range of energy levels (b), rather than when restricting to a single level (a). While this is mathematically sound one may find this physically counterintuitive as single shot work is colloquially often associated with a worst case scenario. The sketch in Fig.~\ref{fig:pic}b intuitively suggests that the worst case result for jumping from the ground state to a range of energy states is a single level transition to the lowest level, implying the lowest energy in (b) is the same as the single energy in (a), $\tw = w$. However, this is not correct - the salient point is that populations of energetic levels higher  than $w$ would not count as ``success'' in the situation depicted in Fig.~\ref{fig:pic}a.  To achieve a high probability of success, $1-\eps$, in  (a) the population needs to be concentrated in just that one level and this is what leads to a lower amount of extractable work in comparison to case (b), i.e. $\tw > w$. 
This new result may appear to contrast with a recent paper \cite{Egloff12} that shows that allowing population in higher levels should not change the ``predicted'' work. However, the work definition used there follows traditional statistical physics concepts, without including a weight explicitly and involves sequences of shifting energy levels in the system Hamiltonian and coupling the system to a bath \cite{Aberg13, AG13}. The work concept thus differs significantly from the  work concept used here.

Now the question is how significant it is if the weight ends up exactly with energy, say $w$, or with an energy in a range, say $[w, w+\delta]$? 
The second case is of relevance for many mechanical processes in physics and engineering where gaining work is the key aim, and obtaining a little more 
energy would be judged positively. There are other situations where \emph{resonance} with a particular energy value, i.e. case (a), is key for performance, 
for instance, this may be the case in biological energy conversion processes, such as photosynthesis \cite{bioref}. We propose that the work ``extracted'' in 
such resonance processes may be described as ``resonance work'' or ``matched work''. To decide which approach to use to calculate work, e.g.  the single shot 
approach with single level transitions (a) or multiple level transitions (b), or other approaches, one has to first identify what is really needed - energy of 
a certain amount or energy above a certain threshold.  

\section{Generalisation of  work extraction to multiple levels in the initial and final state of the  work storage system} \label{sec:casec} \label{sec:multinifi}

\begin{figure}[t]
	\begin{center}
	\includegraphics[width=0.7\textwidth]{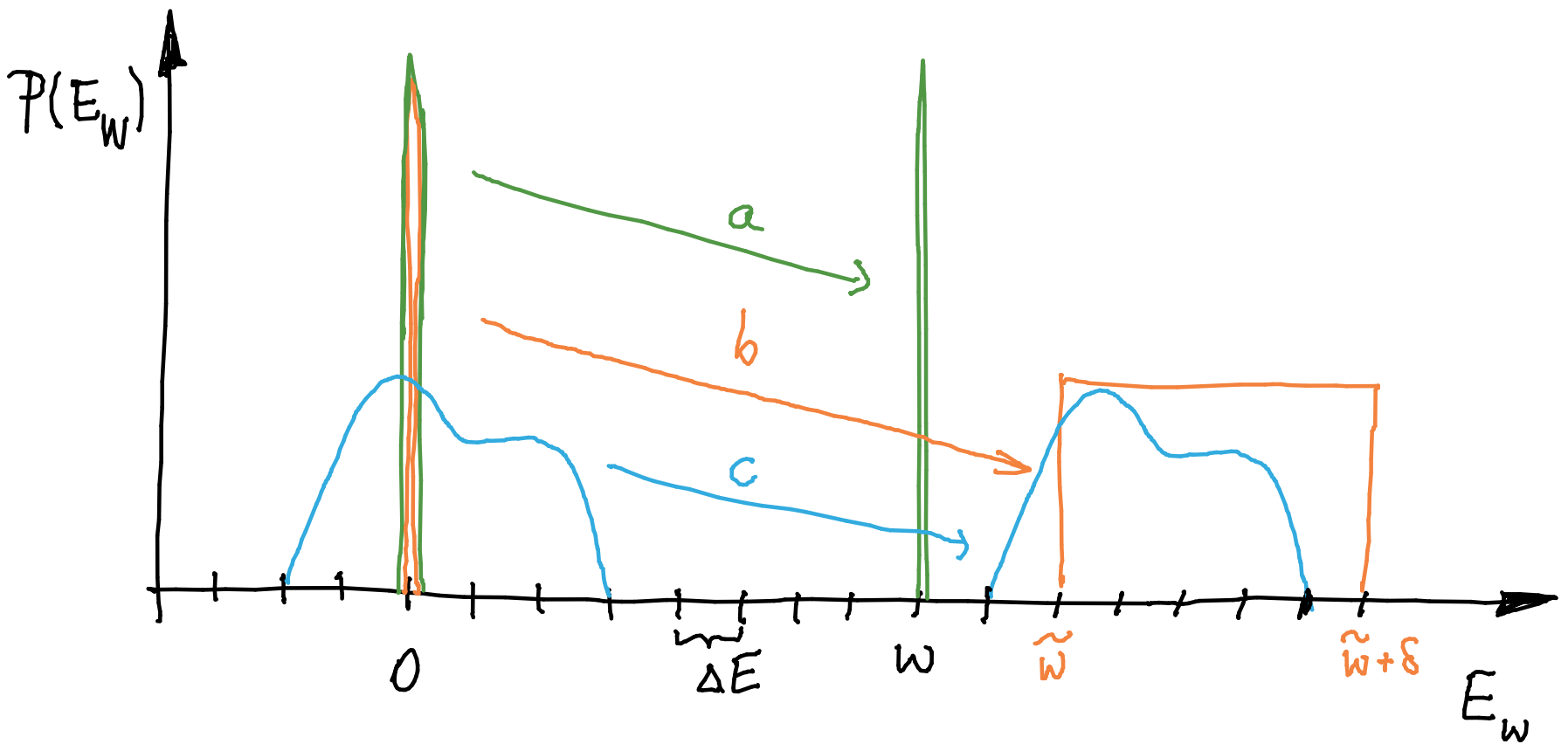}
        \caption{\label{fig:probdist} 
       Weight energy probability distributions, $P(E_W)$, over weight energy, $E_W$. Sketched are 
       (a) (green) a single level to a single level transfer, 
       (b) (orange) a single level to a range of levels transfer, 
       (c) (blue) a range of levels to a range of levels transfer, cf. Fig.~\ref{fig:pic}.
       The particular choice of the distributions is \emph{not generic} - the orange box shows the highest entropy state for that particular range,
       and the initial and final blue distributions are chosen the same shape, implying that these distributions have the same entropy. }
   	\end{center}
\end{figure}

We found that the work definition based on a ``single energy level to single energy level transition'' of a weight, case (a) in Fig.~\ref{fig:pic}, is too narrow and at odds with our notion of classical work storage systems. A physical interpretation was however possible in the context of matching a particular energy, i.e. in  a resonance situation. For the physically motivated extension to multiple final levels, case (b) in Fig.~\ref{fig:pic}, that corresponds to work extraction in the sense of ``at least this much or more energy'' the predictions came out unphysical. So one is left with the unsatisfactory situation that there is no consistent single shot work extraction concept for general initial and final states of the weight. Such extension issues have been recognised by others and motivated a number of recent publications that provide bounds on the single shot work from majorisation \cite{Egloff12} and fluctuation theorems \cite{Halpern14}, address the shifting of general weight probability distributions \cite{Skrzypzyk2014}, and employ axiomatic approaches to identify sets of monotone functions that can act as work quantifiers for specific situations \cite{GEW15, Perry15}. 

To try and rectify this situation we here proceed to allow multiple energy levels in both, the initial and final state, as indicated in Fig.~\ref{fig:pic}c and Fig.~\ref{fig:probdist}c. Here one now faces the problem of defining work on the basis of an initial and a final \emph{energy distribution of the weight}. When the weight transfers from multiple levels to other multiple levels without a means to distinguish any of these transitions, there is no single energy difference $\Delta E_W$ that can be straightforwardly linked to work. Even if one considers the multitude of energy differences associated with the different single level transitions, it is challenging to formulate a ``worst case scenario'' for the reasons discussed in the previous section. We will here not attack the complicated problem of defining single shot work for transfers between multiple levels of the weight, see Fig.~\ref{fig:pic}c and Fig.~\ref{fig:probdist}c, and finding the optimum work. 
Instead, our aim here will be to find a  \emph{transfer-quantity}, which we denote $\savg{w}$, that indicates whether an initial weight energy distribution can or can not be transformed into a final distribution. To analyse the general multiple level scenario we consider again a global unitary ($V$) commuting with the sum of local Hamiltonians as before and no restrictions on their spectra. $V$ maps an initially factorised state to some final possibly correlated state,
\be \label{eq:mapc}
	\eta : = \rho_S \otimes \tau_{B} \otimes \sigma_{W} \stackrel{V}{\longrightarrow}  \eta',
\ee 
where the initial bath state is assumed thermal and the weight's final reduced state is $\sigma'_W := \tr_{SB}[\eta']$. Various transfers between 
different energy distribution of the weight are sketched in Fig.~\ref{fig:probdist}.

The average energy change of the weight on its own cannot be significant to limit thermodynamic transformations. For example, a weight starting in the ground state can have its average energy  raised just by coupling it to a heat bath. Intuitively, the transfer-quantity should reward energy increase while punishing spreading the energy across different energy levels of the weight. It is known that the free energy has exactly such a balancing property and this motivates us to define the transfer-quantity for bringing a weight from state $\sigma_W$ to state $\sigma_W'$ as the free energy difference of the weight,
\be \label{eq:avgw}
    \savg{w}: = \Delta F_W = \Delta U_W - T \Delta S_W.
\ee
Here $\Delta U_W = \tr[H_W (\sigma'_W - \sigma_W)]$ and $\Delta S_W/k_B = - \tr[\sigma'_W \, \ln \sigma'_W]  +  \tr[\sigma_W \, \ln \sigma_W ]$ are the average energy change and entropy change of the weight and $T$ is the temperature of the initial bath state. 
Using sub-additivity of the von Neumann entropy it is straightforward to show, see \ref{sec:statrqu}, that the \emph{transition Eq.~(\ref{eq:mapc}) can only be possible when the transfer-quantity obeys \emph{at least} the following inequality}
\be \label{eq:ineq}
       \savg{w} \leq F (\rho_S) - F(\rho'_S).
\ee
Here $F (\rho_S) - F (\rho'_S) = - \Delta F_S = - \Delta U_S + T \Delta S_S$ is the free energy difference of the two, in general, non-equilibrium system states $\rho_S$ and $\rho'_S = := \tr_{BW}[\eta']$, with $\Delta U_S$ and $\Delta S_S$ defined analogously to above, using instead the system Hamiltonian $H_S$. 
Note, that inequality Eq.~(\ref{eq:ineq}) is necessary but not sufficient, i.e. there are examples where this inequality is fulfilled and the transition may still not be possible. Equality in Eq.~(\ref{eq:ineq}) can only occur when the final global state factorises with the final bath state being a thermal state, see \ref{sec:statrqu}. Thus the bath $B$ must remain in a thermal state under the optimal work extraction process. This clearly hints in the direction of the bound being reachable in the classical macroscopic limit as was suggested in \cite{Horodecki2013}.

Now one can consider special cases of this general relation. A subclass of map Eq.~(\ref{eq:mapc}) are the single level to single level transitions, sketched in Fig.~\ref{fig:pic}a and in Fig.~\ref{fig:probdist}a. The necessary condition for a transformation $\rho_S \to \rho'_S$ to be thermodynamically allowed here becomes
\be \label{eq:casea}
	\savg{w} \stackrel{a}{=}  w \leq F (\rho_S) - F (\rho'_S).
\ee
To maximise $w$ one clearly wants to choose the final state thermal, $\rho'_S =\tau_S$, resulting in the highest permissible value  $w^{\max} = F (\rho_S) - F (\tau_S)$. In general one has $F_{\eps}^{\min} (\rho_S) < F (\rho_S)$, and hence $w^{\max}_{\eps} < w^{\max}$, with $F_{\eps}^{\min} (\rho_S)$ approaching $F(\rho_S)$ in the i.i.d. limit (infinitely many identical copies) \cite{Horodecki2013}. This means that inequality Eq.~(\ref{eq:casea}) is here a lesser requirement than the tight bound Eq.~(\ref{eq:ho}), giving just an upper threshold of what transformations may be allowed.

For weight transitions from a single level to multiple levels, sketched in Fig.~\ref{fig:pic}b, the inequality gives
\be
        \savg{w}
         \stackrel{b}{=}  F (\sigma'_W)   \leq F (\rho_S) - F (\rho'_S).
\ee
So it is down to the choice of a suitable final weight state $\sigma'_W$ to enable the transformation to be thermodynamically possible. Fig.~\ref{fig:probdist}b shows an example of a final energy distribution of the weight (orange box). It can be seen that if one was to shift up the final distribution in energy while keeping the  entropy the same, $F (\sigma'_W)$ would grow thus making the inequality continuously tighter. The limit on how far the final weight distribution can be shifted in energy and the transition still being allowed is given by the  system's free energy difference. 

For a multiple level to multiple level transition, sketched in Fig.~\ref{fig:pic}c,  an example initial weight distribution is sketched in Fig.~\ref{fig:probdist}c. If the energy distribution of the weight is solely shifted to a higher energy without changing its shape or entropy, as indicated in blue in Fig.~\ref{fig:probdist}c,  then the inequality reduces to 
\be
	\savg{w}
	 \stackrel{c}{=}  \Delta U_W 
         \leq F (\rho_S) - F (\rho'_S).
\ee
This result coincides with the \emph{tight} bound derived for a more restrictive process in which global unitaries are chosen to be of a specific form but for which only average global energy conservation is required \cite{Skrzypzyk2014}.

Condition Eq.~(\ref{eq:ineq}) produces sensible results for the special cases discussed here. While for the transformation to be thermodynamically possible fulfilling the condition is only necessary - not sufficient - the condition's strength lies in its applicability for general globally energy conserving transformations on initially factorised states. Apart from providing an easy-to-check criterion to rule out many potential weight transitions, calling the transfer-quantity ``work'' can be supported, see e.g. \cite{GEW15}. 
Indeed we observe that condition Eq.~(\ref{eq:ineq}) for the transfer quantity $\savg{w}$ is of the same form as the well-known bound on the \emph{average work} $\savg{W}$ that can be extracted from a statistical ensemble when changing its probability distribution $\rho_S$ to another probability distribution $\rho_S'$, e.g. in the context of Jarzynski's equality \cite{Jarzynski}, 
\be	
	       \savg{W} \leq - \Delta F_S,
\ee
which is widely viewed as the second law of thermodynamics. 

\section{Discussion} \label{sec:discussion}

We have re-derived the extractable work and the work of formation in the single shot setting proposed in \cite{Horodecki2013}, where a work storage system (a) transfers from a single energy level to another single energy level with probability $1-\eps$. Our alternative derivation of the optimal work value for which such a transformation is still possible was based on the comparison of subspace dimensions while no discussion of thermo-majorisation was required. (Thermo-majorisation  \cite{Horodecki2013} offers an additional and separate method of addressing the same optimisation problem.)
The approach presented here facilitates a discussion of the final state of the system after work extraction has been performed. Indeed, we find that \emph{typically} the reduced final state is a minimum free energy thermal state as has been suggested \cite{Horodecki2013}. ``Typically'' is here to be understood in the very same sense in which it is often used in the context of thermalisation studies: it is overwhelmingly frequent w.r.t. Haar distributed unitaries \cite{Reimann2007, Goldstein2006, Popescu2006, Gemmer2003}. The implication of this finding is two-fold: on one hand it should not come as a surprise if a ``work-extracted'' system is left in a thermal state, on the other hand the question of how far away from a thermal state a system may end up for a specific, for instance physically motivated, unitary is a promising direction for future research.

The presented approach opened the possibility to discuss a single shot work concept when transitions of the work storage system between (b) a single energy level and \emph{a range of final energy levels} is permitted. Extensions to multiple levels are desirable to reflect experimental constraints - current and near-future experimental control of meso/macro-scopic systems does not allow to distinguish between energy levels that are spaced by less than an optical phonon. While the results for the multiple final level situation are mathematically sound, they raise interpretation issues when applied to the physical context. Besides showing an apparent violation of the second law that requires careful attention, an important conclusion emerges. 
The term ``work extraction'' intuitively suggests that an energy of \emph{a certain amount or higher} is being made available but the ``minimum extractable work'' found for case (b) was in fact higher than the single level case (a) and thus at odds with physical intuition. The single shot work analysis thus turns out to be not a good setting for discussing ``work extraction''. It is however rather suited to characterise \emph{resonance processes} where a specific amount of energy needs to be stored in the work storage system with high probability. Due to its nature, the energy stored may here be thought of as \emph{resonance work}. 

Finally, the aim was to consider multiple energy levels of the work storage system for both, initial and final state. We have not tried to generalise the single shot work extraction approach to this situation because of the issues in physical interpretation that we could not fully resolve. Instead we provide a means to decide whether a transfer can be thermodynamically allowed or not by introducing a transfer-quantity $\savg{w}$ that we chose as the \emph{free energy of the weight}. The derived necessary bound, $\savg{w} \le - \Delta F_S$, is applicable for transformations with general initial and final states of the weight. The bound is not tight in general, i.e. there are transformations that fulfil the bound while still not being possible with the thermodynamic resources considered here. However, the bound gives tight results for the single level case (a) when applied to $N \to \infty$ independent identical (i.i.d) copies \cite{Horodecki2013,Aberg13} and to a set of restricted global unitaries that results in multi-level to multi-level weight transfers \cite{Skrzypzyk2014}. Since the criterion is fairly simple to apply it facilitates the discussion of many practical limits on work extraction in the quantum regime. To rephrase, while Eq.~(\ref{eq:ineq}) is not precisely tight it is sufficiently tight to be relevant and thus it may be considered a valuable concept.

\ack 
The authors thank the Institute for Mathematical Science (IMS) in Singapore for the stimulating environment at the ``Information-Theoretic Approaches to Thermodynamics'' meeting at which this project was started. This work was supported by the European COST network MP1209. JA acknowledges support by the Royal Society and EPSRC (EP/M009165/1) in the UK. 


\appendix

\section{Derivation of properties of the final states } \label{derfin}

This part of the Appendix is dedicated to a more detailed derivation of the results on the final states discussed in Section \ref{sec:fisysta}.

We start by denoting the projector spanned by all eigenstates of the initial state $\eta $ that have total energy $E$ and $h(E_S, g)=1$ by $\sini^E(\eps, \eta)$. The dimension of this projector  $d^E_{\ini}(\epsilon, \eta)= \tr[ \sini^E(\eps, \eta) ]$ has already been calculated in (\ref{eq:dinie}). By construction it equals the dimension $d^E_{\fin}$ of the respective projector $\sfin^E$.  The maximum work extraction itself simply consists in a unitary $V$ which maps the initial subspaces on the respective final ones, thereby respecting energy shells, 
which is the only condition on $V$. 
As mentioned in Sect. \ref{sec:fisysta} the crucial point here is that there is obviously not a single, unique unitary that corresponds to maximum work extraction but infinitely many. Since there are very many unitaries there may be potentially very many different reduced states of the system after maximum work extraction. In  order to better understand the multitude of final system states it is instructive to consider unitaries which are randomly distributed according to the Haar measure. (Those unitaries may be constructed by drawing (column) vectors component-wise at random and doing a Gram-Schmidt orthonormalization For more details on averaging over Haar distributed unitaries see \cite{Goldstein2010}). 
In the following we outline some results on distribution of quantities generated by such random unitaries without (formal) proof. 

Consider the unitary mapping $V$ of some matrix $A$ and a subsequent projection onto its diagonal part w.r.t. some orthonormal basis $\{|n \rangle\}$, i.e., $\sum_n|n \rangle\ \!\! \langle n|VAV^{\dagger}|n \rangle\ \!\! \langle n|$. The average over all such unitaries may simply be inferred from symmetry considerations:
\be \label{units}
 	\avg{\sum_{n=1}^d|n \rangle\ \!\!  \langle n|VAV^{\dagger}|n \rangle\  \!\! \langle n|} 
	=\frac{\tr [A]}{d} \, \hat{\Pi},  
 \quad \hat{\Pi}:=\sum_{n=1}^d|n \rangle\ \!\! \langle n |
\ee
where $\avg{\cdot}$ denotes the average over all unitaries and $d$ is the dimension of the Hilbertspace, $d= \tr[\hat{\Pi}]$.
Denoting  the sum over $m (\leq d)$ of the diagonal elements of this operator by $s$, we find for the latter:
\be \label{defsum}
 	s:=\sum_{n=1}^m \langle n|VAV^{\dagger}|n \rangle, 
\ee
we find from (\ref{units}) for the average of $s$:
\be \label{typav}
 	\avg{ s } =\frac{m \, \tr[A]}{d}.
\ee
It is more subtle to compute  the variance of $s$ but it has been done, e.g.,  in the context of  quantum typicality \cite{Goldstein2006,Gemmer2003}. From the latter works we find:
\be \label{typvar}
 	(\Delta s)^2:=\avg{ s^2 } -   \avg{ s }^2  =\frac{m(1-\frac{m}{d}) \, \tr[A]^2}{d(d+1)},
\ee
and for the relative deviation (setting $d+1 \approx d$ for simplicity)
\be \label{reldev}
 	\frac{\Delta s}{\avg{ s }}   :=\avg{ s^2 } -  \avg{ s }^2  =\sqrt{\frac{1}{m}(1-\frac{m}{d})}.
\ee
Thus, if $m$ is sufficiently large, those relative deviations become (negligibly) small. Which means even for a single unitary drawn at random according to the Haar measure $s$ may be expected to be given by $s \approx \avg{ s }$ for the overwhelming majority of cases.

Equipped with these findings we may now identify the ``typical'' local reduced final state of the system $S$. To this end we consider (\ref{units}) and make the identifications: 
\be \label{ascrp}
 	A: \sini^E(\eps, \eta) \, \eta \, \sini^E(\eps, \eta), \quad  \quad  \hat{\Pi}: \sfin^E, 
\ee
furthermore we note that
\be \label{probe}
	\tr[A] = \tr[\sini^E(\eps, \eta) \, \eta \, \sini^E(\eps, \eta)] = (1-\epsilon) \,  P^E_{\ini}.
\ee
Let $\eta'_{\diag}$ be the diagonal part (w.r.t. the product energy eigenstates of the uncoupled subsystems) of the full system final state that has actually changed under work extraction, i.e. everything except for the $\epsilon$-part that remained unchanged, and  $\eta'_{\diag,E}$ the state's projection into the subspace of energy $E$. Then one finds from (\ref{units}), inserting (\ref{ascrp}) and (\ref{probe}):
\be \label{fina}
	\avg{ \eta'_{\diag,E}}=   \frac{(1-\epsilon) \, P^E_{\ini} \, \sfin^E}{d_{\fin}^E}.
\ee
Since $P^E_{\ini}$ is invariant under $V$ it may be calculated most conveniently from (\ref{eq:roini}). From this calculation it is found to be actually independent of $E$, i.e., the total probabilities on all energy shells are the same. Using (\ref{eq:pfine}) and (\ref{eq:dfine}) yields
\be \label{finb}
	\avg{ \eta'_{\diag,E}}\propto \frac{1}{e^{\beta E}}\sum_{E_S}\hat{\Pi}_S(E_S)\otimes \hat{\Pi}_B(E-E_S-w) \otimes \hat{\Pi}_W(w).
\ee
At this point it is actually more convenient to go back to the original, local energies and sum over $E_S, E_B$ rather than $E_S, E$. Doing so yields:
\be \label{finc}
	\avg{ \eta'_{\diag}}\propto\sum_{E_S, E_B}\hat{\Pi}_S(E_S)e^{-\beta E_S}\otimes \hat{\Pi}_B(E_B-w)e^{-\beta E_B} \otimes \hat{\Pi}_W(w)
\ee
Using the index shift $E_B-w \rightarrow E_B$, this may be summed as 
\be \label{find}
	\avg{ \eta'_{\diag}}\propto e^{-\beta \hat{H}_S}\otimes  e^{-\beta \hat{H}_B}   \otimes \hat{\Pi}_W(w)
\ee
which is (\ref{eq:find}) as given and discussed in Section \ref{sec:fisysta}.

However, to conclude  that the actual final state of the system $S$ corresponding to a single random unitary is close to $ e^{-\beta \hat{H}_S}/Z_S$ and thus a minimum free energy state it remains to be shown that the relative deviations of $\eta'_{\diag}$ are small. To this end it is instructive to write out the diagonal matrix elements of the reduced final system state $\sigma_S = \tr_{BW}[\eta']$ explicitly  
\be \label{trace}
	&&\langle E_S, g| \sigma_S| E_S, g \rangle = \sum_E P_{\fin}(E,  E_S, g),\\
	&& P_{\fin}(E,  E_S, g):=  \sum_{f=1}^{M_B(E_B)}\langle E_S, g, E, f,w| \eta'_{\diag}| E_S, g, E, f,w \rangle, \nonumber
\ee
where $| E_S, g, E, f, w \rangle$ are the respective (product) energy eigenstates of the decoupled system. So a single  $P_{\fin}(E,  E_S, g)$, i.e., a contribution to the respective diagonal element of the final reduced state of $S$ from some energy shell $E$, is itself a sum over very many diagonal elements of the full final state, as may be seen from the lower line of (\ref{trace}). Namely, it  runs over all bath energy eigenstates at  $E_B= E-E_S-w$, i.e., contains $M_B(E_B)$ summands. Thus from (\ref{reldev}), setting $m =M_B(E-E_S-w)$, it may be inferred that the relative deviations of the contributions to the diagonal matrix element are on the order of
\be \label{reldevdiag}
 	\frac{\Delta  P_{\fin}(E,  E_S, g)  }{\avg{ P_{\fin}(E,  E_S, g) }}
	 \approx \frac{1 }{\sqrt{M_B(E-E_S-w) }},
\ee
which is very small if the bath is large. This is (\ref{eq:fluc}) as given and discussed in Section \ref{sec:fisysta}. Furthermore,  if the unitaries are drawn at random, the $\hat{U}_E$ are independent of each other. Thus one may conclude from (\ref{reldevdiag}) and the central limit theorem, that according to the first  line of (\ref{trace}), 
the relative  deviations of  $\langle E_S, g| \sigma_S| E_S, g \rangle$ will be even smaller if many energy shells are involved.

An analogous consideration (which we do not present here in full detail for brevity) applies to the off-diagonal elements of the final reduced state of $S$. The analogue to (\ref{typav}) required to find the average of those  off-diagonal elements reads \cite{Gemmer2009}
\be \label{typavoff}
 	\avg{ \langle n|VAV^{\dagger}|n' \rangle } =0  \quad \mbox{for} \quad n \neq n'.
\ee
Thus, using the same identifications as before (\ref{ascrp}), on average off-diagonal elements of the full system final state vanish. Since off-diagonal elements of reduced states are sums of off-diagonal elements of the full state exclusively, the averages of the final reduced state of the system $\sigma_S$ also vanish. Thus on average, the final state is indeed diagonal in its energy eigenbasis. Furthermore only those (off-diagonal) elements of the full state for which the full system states $|n' \rangle, |n \rangle$ both correspond to the same bath system state  $|E_B,f \rangle $ contribute to the reduced state of $S$ at all. Thus each contribution to the reduced state of  $S$ may be associated with a certain energy shell $E_B$ of the bath. An  analogue to (\ref{typvar}) quantifies the deviations of off-diagonal elements, 
\be \label{typavoff1}
	(\Delta o)^2:=\avg{ \, |\langle n|VAV^{\dagger}|n' \rangle|^2 \, } \quad \mbox{for} \quad n \neq n', 
\ee
and reads, according to e.g. \cite{Gemmer2009}, 
\be \label{typavoff2}
 	(\Delta o)^2:=\frac{\tr [A^2] }{d(d+1)}.
\ee
From (\ref{ascrp}) and (\ref{units}) it may be inferred that, $\tr[ A^2 ] \leq \tr [\sini^E(\eps, \eta)]$ and $d= \tr[ \sfin^E]$. In Sect. \ref{sec:disc} is has been established that both traces essentially scale with the dimension of the bath, i.e. 
\be \label{scale}
 	\tr[ \sini^E(\eps, \eta) ] \propto M_B(E), \quad \tr[ \sfin^E ] \propto M_B(E).
\ee
Thus we find 
\be \label{typavoff3}
 	\Delta o \leq \frac{C}{\sqrt{M_B(E)}}.
\ee
where $C$ is a number which  is independent of the overall dimension of the bath. As $M_B(E) \geq  {M_B(E-E_S-w)}$, this implies that the deviations of the off-diagonal elements of the reduced state of $S$ again scale at most as $\frac{1 }{\sqrt{M_B(E-E_S-w)}}$, just like in the case of the diagonal elements, cf. (\ref{reldevdiag}).  This result is also given and discussed in Section \ref{sec:fisysta}.

\section{Derivation of  upper bound to the state transfer quantity $\langle w \rangle$}
\label{sec:statrqu}
This part of the appendix is dedicated to the derivation of (\ref{eq:ineq}) in Sect. \ref{sec:multinifi}. In the remainder any state of the total system, possibly including correlations,  will be denoted by $\eta$. Reduced local states will  be denoted by $\rho_{S}= \tr_{BW}[\eta]$ for the non-equilibrium system, $\sigma_{W}=  \tr_{SB}[\eta]$ for the weight storage system, and  $\tau _{B}= \tr_{SW}[\eta]$ for the bath. Just as before $V$ transforms the total initial state  $\eta$ into the total final state $\eta'$ which means $\eta'= V\eta V^{\dagger} $. And just as before we assume conservation of the sum of local energies, i.e., $[H_S+H_B+H_W, V]=0$.
Free energies will be denoted as 
\be \label{freeen}
	F(X):=U(X)  -\frac{S(X)}{\beta} \quad \mbox{with} \quad S(X):=- \tr [ X \, \ln X],
\ee
where $X:=\rho_{S}, \sigma_{W}, \tau_B, \eta$. Thus, free energies may refer either to parts or to the total system. The expectation value of the respective energy is denoted by $U(X)$ for the corresponding Hamiltonian, $H_S, H_W, H_B$. Generally primed operators will refer to final states and unprimed operators to initial states. 

Again, as also mentioned in Sect. \ref{sec:multinifi}, we assume factorizing initial conditions, i.e., $\eta =\rho_S \otimes \tau _{B}\otimes \sigma_{W}$. We also make the assumption of the initial state of the bath is thermal, $\tau_B \propto e^{- \beta H_B}$. The only difference in the set up of the approach in this Section compared to Sections \ref{sec:max}-\ref{sec:disc} is that we do not require the state density of the spectrum of the bath to be exponentially growing. By construction, the total entropy and the sum of local energies are both invariant under $V$.  Thus, the initial  total free energy and the final total free energy are the same, $F(\eta)=F(\eta')$, and due to the factorizing initial conditions we have
\be \label{initotloc}
	F(\eta) =  F(\rho_S)+F(\tau_B)+F(\sigma_W).
\ee
From the Araki-Lieb theorem \cite{Araki-Lieb} it follows that 
\be \label{al}
	S(\eta')\leq S(\rho'_S)+S(\tau'_{B})+S(\sigma'_{W})
\ee
and hence, 
\be \label{fineq}
	F (\eta')\geq F(\rho'_S)+F(\tau'_{B})+F(\sigma'_{W}).
\ee
Combining (\ref{initotloc}) and (\ref{fineq}) yields
\be \label{ifineqf}
 	F(\rho_S)+F(\tau_B)+F(\sigma_W)    \geq F(\rho'_S)+F(\tau'_{B})+F(\sigma'_{W})
\ee
which may simply be rearranged as
\be \label{rear}
 	F(\sigma'_{W})- F(\sigma_W)       \leq -( F(\rho'_S)- F(\rho_S)   )-( F(\tau'_{B})- F(\tau_B) ).
\ee

A thermal state has the lowest free energy given a specific inverse temperature $\beta$. Since we assumed an initial thermal state for the bath it follows that $F(\tau_B) $ is the lowest possible bath free energy. Consequently, whatever $F(\tau'_{B})$ is, one has 
\be \label{ineqfreebath}
	-(F(\tau'_{B})- F(\tau_B)  ) \leq 0.
\ee
By dropping the corresponding term in  (\ref{rear}) we only make the r.h.s. possibly larger. Since this is not in conflict with the inequality in (\ref{ineqfreebath}) we may simply drop the term, obtaining
\be \label{diffree}
	F(\sigma'_{W})- F(\sigma_W)   \leq -( F(\rho'_S)- F(\rho_S)   ).
\ee
The l.h.s. is $F(\sigma'_{W})- F(\sigma_W)= \langle w \rangle$ as defined in (\ref{eq:avgw}) completing  the derivation of (\ref{eq:ineq}).


\section*{References}

\begin{thebibliography}{99}

\bibitem{Horodecki2013}
Horodecki, M. \& Oppenheim, J.
{\it Nat. Commun.} {\bf 4,} 2059 (2013).

\bibitem{Evans93}
Evans, D.J., Cohen, E.G., \&  Morriss, G.P.
{\it Phys. Rev. Lett.} {\bf 71}, 2401 (1993). 

\bibitem{Jarzynski}
Jarzynski, C.
{\it J. Stat. Phys.} {\bf 96}, 415 (1999).

\bibitem{Crooks}
Crooks, G.
{\it Phys. Rev. E} {\bf 60}, 2721 (1999).

\bibitem{Tasaki}
Tasaki, H.
arXiv:cond-mat/0009244v2 (2000).

\bibitem{KURCHAN}
Kurchan, J.
arXiv:cond-mat/0007360v2 (2000).

\bibitem{Mukamel}
Mukamel, S. 
{\it Phys. Rev. Lett.} {\bf 90}, 170604 (2003).

\bibitem{Kawai}
Kawai, R., Parrondo, J.M., \& Van den Broeck, C.
{\it Phys. Rev. Lett. } {\bf 98}, 1 (2007).

\bibitem{Hanggi}
Talkner, P., Lutz, E., \& H\"anggi,  P. 
{\it Phys. Rev. E (R)} {\bf 75}, 050102 (2007).

\bibitem{CAMPISI2011}
Campisi, M., H\"anggi,  P., \& Talkner, P. 
{\it Rev. Mod. Phys.} {\bf 83}, 771 (2011). 

\bibitem{Paternostro}
Mazzola, L.,  De Chiara, G., \& Paternostro, M.
{\it Phys. Rev. Lett.} {\bf 110}, 230602 (2013).

\bibitem{Paz}
Roncaglia, A.J., Cerisola, F., \& Paz, J.P.
{\it Phys. Rev. Lett.} {\bf 113}, 250601 (2014).

\bibitem{Janzing}
Janzing, D., Wocjan, P., Zeier, R., Geiss, R. \& Beth, T. 
{\it Int. J. Theor. Phys.} {\bf 39}, 2717 (2000).

\bibitem{resource}
Brandao, F. G. S. L., Horodecki, M., Oppenheim, J., Renes, J. M. \& Spekkens, R. W. 
{\it Phys. Rev. Lett.} {\bf 111}, 250404 (2013). 


\bibitem{Aberg13}
{\AA}berg, J.
{\it Nat. Commun.} {\bf 4,} 1925 (2013).

\bibitem{SSP14}
Skrzypzyk, P., Short, A.J. \&  Popescu, S.
{\it Nat. Commun.} {\bf 4,} 4185 (2013).

\bibitem{Skrzypzyk2014}
Skrzypzyk, P., Short, A.J. \&  Pospescu, S.
{\it Nat. Commun.} {\bf 5,} 4185 (2014).

\bibitem{Brandao13b}
Brandao, F. G. S. L., Horodecki, M., Ng, N. H. Y., Oppenheim, J. \& Wehner, S. 
{\it PNAS} {\bf 112}, 3275 (2015). 

\bibitem{Lostaglio15}
Lostaglio, M., Jennings, D., \& Rudolph, T.
{\it Nat. Commun.} {\bf 6,} 6383 (2015).

\bibitem{KA15}
Kammerlander, P., \& Anders, J.
{\it arXiv:1502.02673 [quant-ph]} (2015).

\bibitem{Gemmer2009}
Gemmer, J., Michel,  M. \& Mahler, G.
\emph{Quantum Thermodynamics}
Springer  (2009).

\bibitem{vonNeumann}
Birkhoff, D.
{\it Univ. Nac. Tucuman Rev. Ser. A} {\bf 5}, 147 (1946).

\bibitem{Ando}
Ando, T.
{\it Linear Algebra and its Applications} {\bf 118}, 163 (1989). 

\bibitem{Mat}
Bhatia, R.
{\it Matrix Analysis} Springer (1996)

\bibitem{Ref1}
Nielsen, M.A. \& Vidal, G. 
{\it Quantum Inf. Comput.} {\bf 1}, 76 (2001).

\bibitem{Ref2}
Mari, A., Giovannetti, V, \& Holevo, A.S.
{\it Nature Commun.} {\bf 5} 3826 (2014).


\bibitem{Goldstein2010}
Goldstein, S., Lebowitz, J.L., Mastrodonato, C., Tumulka, R. \&  Zanghi, N.
{\it Eur. Phys. J. B} {\bf 31,} 249 (2003).

\bibitem{Reimann2007}
Reimann, P.
{\it Phys. Rev. Lett.} {\bf 99,} 160404 (2007).

\bibitem{Goldstein2006}
Goldstein, S., Lebowitz, J.L., Tumulka, R. \&  Zanghi, N.
{\it Phys. Rev. Lett.} {\bf 96,} 050403 (2006).

\bibitem{Popescu2006}
Popescu, S., Short, A.J. \& Winter, A.
{\it Nature Physics } {\bf 2,} 754 (2006).

\bibitem{Gemmer2003}
Gemmer, J. \& Mahler, G.
{\it Eur. Phys. J. B} {\bf 31,} 249 (2003).

\bibitem{Lubkin1978}
Lubkin, E.
{\it J. Math. Phys.} {\bf 19,} 1028 (1978).

\bibitem{Page1993}
Page, D.N.
{\it Phys. Rev. Lett.} {\bf 71,} 1291 (1993).

\bibitem{Lloyd1988}
Lloyd, S. \&  Pagels, H.
{\it Ann. Phys. (N.Y.)} {\bf 188,} 186 (1988).

\bibitem{Sen1996}
Sen, S.
{\it Phys. Rev. Lett.} {\bf 77,} 1 (1996).

\bibitem{Gemmer2001}
Gemmer, J., Otte, A. \& Mahler, G.
{\it Phys. Rev. Lett.} {\bf 86,} 1927 (2001).

\bibitem{Marti14}
Perarnau-Llobet, M., {\it et al.}
{\it arXiv:1407.7765v2} (2014).

\bibitem{HSAL11} 
Hilt, S., Shabbir,  S.,  Anders,  J. \& Lutz,  E.
{\it Phys. Rev. E} {\bf 83,} 030102 (2011). 

\bibitem{Esposito10}
Esposito, M., Lindenberg, K. \& Van den Broeck, Ch.
{\it New J. Phys.} {\bf 12,} 013013 (2010).


\bibitem{Egloff12}
Egloff, D., Dahlsten, O.C.O., Renner, R., \& Vedral, V.
arxiv:1207.0434 (2012).

\bibitem{AG13}
Anders, J. \& Giovanetti, V.
{\it New J. Phys.} {\bf 15,} 033022 (2013).

\bibitem{bioref}
Olaya-Castro, A., Nazir, N., \& Fleming, G.R.
{\it Phil. Trans. R. Soc. A} {\bf 370} 3613 (2012).  

\bibitem{Halpern14}
Yunger-Halpern, N., Garner, A.J.P. , Dahlsten, O.C.O. \& Vedral,  V.
arXiv:1409.3878 (2014).

\bibitem{GEW15}
Gallego, R., Eisert, J., \& Wilming, H. 
arxiv: 1504.05056  (2015).

\bibitem{Perry15}
Alhambra, A.M.,  Oppenheim, J. \& Perry, C. 
arxiv:1504.00020 (2015).



%
%

\bibitem{Araki-Lieb}
Araki, H., \& Lieb, E.H.
{\it Comm. Math. Phys.} {\bf 18} 160  (1970).

\end{thebibliography}
\end{document}